\documentclass{svproc}

\usepackage{lineno, xcolor}

\usepackage{amssymb, amsmath}
\usepackage{tikz}
\usepackage{graphicx, subcaption}
\usepackage{url}
\usepackage{cite}

\begin{document}
\mainmatter              
\title{A New Hybrid Approach for Identifying Obsolescence Features: Applied to Railway Signaling Infrastructure}
\titlerunning{A Hybrid Approach for Identifying Obsolescence Features}  
%
\author{Elie Saad\inst{1,2} \and Mariem Besbes\inst{1} \and Vincent Bourgeois\inst{2} \and Victor Czmil\inst{2} \and Marc Zolghadri\inst{1}}
\authorrunning{Saad et al.} 
%
\tocauthor{Elie Saad, Mariem Besbes, Vincent Bourgeois, Victor Czmil, Marc Zolghadri}
\institute{Institut Sup\'{e}rieur de M\'{e}canique de Paris, Laboratoire Quartz (EA7393), 3 Rue Fernand Hainaut, 93400 Saint-Ouen-sur-Seine, France,\\
\email{\{elie.saad, mariem.besbes, marc.zolghadri\}@isae-supmeca.fr},\\
\and
SNCF R\'{e}seau, D\'{e}partement de Signalisation Ferroviaire, 6 Av. Fran\c{c}ois Mitterrand, 93210 Saint-Denis, France,\\
\email{\{elie.saad, vincent.bourgeois1, victor.czmil\}@reseau.sncf.fr}}

\maketitle              

\begin{abstract}
Electrical component obsolescence poses a major issue especially within systems with large life cycles. Thus, finding the optimal management solution for each obsolescence case is as crucial as knowing what to consider when faced with an obsolescence case. In this paper, a novel hybrid approach for identifying features affecting electrical component obsolescence management is introduced, which combines features engineering techniques and expert knowledge. The method then uses machine learning to predict obsolescence resolution techniques in order to find the optimal resolution. The motivation behind this research is driven by the imperative need for SNCF RESEAU to optimally address and mitigate the challenges posed by electrical component obsolescence in railway infrastructure. 
\end{abstract}
\section{Introduction}
Obsolescence, pertains to the process of becoming non-available, non-adequate, and/or non-suitable~\cite{zolghadri2021obsolescence}. Obsolescence poses a problem as it renders previously functional components obsolete, leading to waste, inefficiency, and the need for constant adaptation. One of the major causes of obsolescence is the difference between the component life cycle ($\approx 18$ months or less) and the systems within which it is contained (30 years or more)~\cite{renaudmacup}, e.g., planes, trains, railway infrastructures, etc.

The difference between the component and system life-cycles triggers component obsolescence that requires a response, sometimes incurring costs rarely acceptable. For instance, the decades-old railway infrastructure of SNCF RESEAU constantly grapples with issues related to the obsolescence of electrical components, resulting in significant and unplanned annual expenditures. For that reason, multiple approaches have been proposed for obsolescence management strategy selection focused on minimizing the cost~\cite{porter1998economic, zheng2015design, singh2006obsolescence}. But various features beyond cost, including the number of suppliers, time constraints, solution reliability, and multiple stakeholder opinions from design, operations, manufacturing, sales, and service, can significantly influence the determination of the optimal decision for obsolescence management~\cite{pingle2015selection}.

The contribution of this work is twofold: proposing an approach for deriving the set of features affecting the decision-making process for obsolescence management, and proposing an analysis method for determining the contribution of each feature within the decision-making process through the use of machine learning. The remainder of this paper is separated as follows: starting with the dataset formation and methodology description in Section \ref{sec:contribution}, followed by the presentation of the results in Section \ref{sec:experimentation}, and finally ending with a brief conclusion and discussion in Section \ref{sec:conclusion_and_perspective}.
\section{Contribution}\label{sec:contribution}
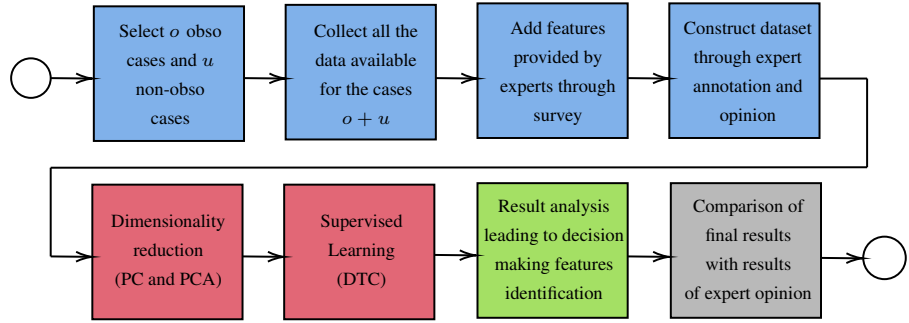
\begin{figure}
    \centering
    \tikzset{every picture/.style={line width=0.75pt}} 
\begin{tikzpicture}[font=\footnotesize, x=0.75pt, y=0.75pt, yscale=-0.75, xscale=0.75]

\draw  [fill={rgb, 255:red, 74; green, 144; blue, 226 }  ,fill opacity=0.7 ] (57,2) -- (158,2) -- (158,92.6) -- (57,92.6) -- cycle ;
\draw   (1.4,50.3) .. controls (1.4,42.95) and (7.35,37) .. (14.7,37) .. controls (22.05,37) and (28,42.95) .. (28,50.3) .. controls (28,57.65) and (22.05,63.6) .. (14.7,63.6) .. controls (7.35,63.6) and (1.4,57.65) .. (1.4,50.3) -- cycle ;
\draw   (574.4,171.3) .. controls (574.4,163.4) and (580.8,157) .. (588.7,157) .. controls (596.6,157) and (603,163.4) .. (603,171.3) .. controls (603,179.2) and (596.6,185.6) .. (588.7,185.6) .. controls (580.8,185.6) and (574.4,179.2) .. (574.4,171.3) -- cycle ;
\draw  [fill={rgb, 255:red, 74; green, 144; blue, 226 }  ,fill opacity=0.7 ] (186,1) -- (287,1) -- (287,91.6) -- (186,91.6) -- cycle ;
\draw  [fill={rgb, 255:red, 74; green, 144; blue, 226 }  ,fill opacity=0.7 ] (315,0) -- (416,0) -- (416,90.6) -- (315,90.6) -- cycle ;
\draw  [fill={rgb, 255:red, 74; green, 144; blue, 226 }  ,fill opacity=0.7 ] (444,0) -- (545,0) -- (545,90.6) -- (444,90.6) -- cycle ;
\draw  [fill={rgb, 255:red, 208; green, 2; blue, 27 }  ,fill opacity=0.6 ] (56,122) -- (157,122) -- (157,212.6) -- (56,212.6) -- cycle ;
\draw  [fill={rgb, 255:red, 208; green, 2; blue, 27 }  ,fill opacity=0.6 ] (185,122) -- (286,122) -- (286,212.6) -- (185,212.6) -- cycle ;
\draw  [fill={rgb, 255:red, 126; green, 211; blue, 33 }  ,fill opacity=0.7 ] (315,121) -- (416,121) -- (416,211.6) -- (315,211.6) -- cycle ;
\draw  [fill={rgb, 255:red, 155; green, 155; blue, 155 }  ,fill opacity=0.7 ] (445,121) -- (546,121) -- (546,211.6) -- (445,211.6) -- cycle ;
\draw    (28,50.3) -- (54,50.58) ;
\draw [shift={(56,50.6)}, rotate = 180.61] [color={rgb, 255:red, 0; green, 0; blue, 0 }  ][line width=0.75]    (10.93,-3.29) .. controls (6.95,-1.4) and (3.31,-0.3) .. (0,0) .. controls (3.31,0.3) and (6.95,1.4) .. (10.93,3.29)   ;
\draw    (158,50.3) -- (184,50.58) ;
\draw [shift={(186,50.6)}, rotate = 180.61] [color={rgb, 255:red, 0; green, 0; blue, 0 }  ][line width=0.75]    (10.93,-3.29) .. controls (6.95,-1.4) and (3.31,-0.3) .. (0,0) .. controls (3.31,0.3) and (6.95,1.4) .. (10.93,3.29)   ;
\draw    (287,50.3) -- (313,50.58) ;
\draw [shift={(315,50.6)}, rotate = 180.61] [color={rgb, 255:red, 0; green, 0; blue, 0 }  ][line width=0.75]    (10.93,-3.29) .. controls (6.95,-1.4) and (3.31,-0.3) .. (0,0) .. controls (3.31,0.3) and (6.95,1.4) .. (10.93,3.29)   ;
\draw    (416,50.3) -- (442,50.58) ;
\draw [shift={(444,50.6)}, rotate = 180.61] [color={rgb, 255:red, 0; green, 0; blue, 0 }  ][line width=0.75]    (10.93,-3.29) .. controls (6.95,-1.4) and (3.31,-0.3) .. (0,0) .. controls (3.31,0.3) and (6.95,1.4) .. (10.93,3.29)   ;
\draw    (157,170.3) -- (183,170.58) ;
\draw [shift={(185,170.6)}, rotate = 180.61] [color={rgb, 255:red, 0; green, 0; blue, 0 }  ][line width=0.75]    (10.93,-3.29) .. controls (6.95,-1.4) and (3.31,-0.3) .. (0,0) .. controls (3.31,0.3) and (6.95,1.4) .. (10.93,3.29)   ;
\draw    (286,169.3) -- (312,169.58) ;
\draw [shift={(314,169.6)}, rotate = 180.61] [color={rgb, 255:red, 0; green, 0; blue, 0 }  ][line width=0.75]    (10.93,-3.29) .. controls (6.95,-1.4) and (3.31,-0.3) .. (0,0) .. controls (3.31,0.3) and (6.95,1.4) .. (10.93,3.29)   ;
\draw    (416,170.3) -- (442,170.58) ;
\draw [shift={(444,170.6)}, rotate = 180.61] [color={rgb, 255:red, 0; green, 0; blue, 0 }  ][line width=0.75]    (10.93,-3.29) .. controls (6.95,-1.4) and (3.31,-0.3) .. (0,0) .. controls (3.31,0.3) and (6.95,1.4) .. (10.93,3.29)   ;
\draw    (546,171.3) -- (572,171.58) ;
\draw [shift={(574,171.6)}, rotate = 180.61] [color={rgb, 255:red, 0; green, 0; blue, 0 }  ][line width=0.75]    (10.93,-3.29) .. controls (6.95,-1.4) and (3.31,-0.3) .. (0,0) .. controls (3.31,0.3) and (6.95,1.4) .. (10.93,3.29)   ;
\draw    (28,170.3) -- (54,170.58) ;
\draw [shift={(56,170.6)}, rotate = 180.61] [color={rgb, 255:red, 0; green, 0; blue, 0 }  ][line width=0.75]    (10.93,-3.29) .. controls (6.95,-1.4) and (3.31,-0.3) .. (0,0) .. controls (3.31,0.3) and (6.95,1.4) .. (10.93,3.29)   ;
\draw    (28,110.6) -- (28,170.3) ;
\draw    (28,110.6) -- (577,110.3) ;
\draw    (545,50.6) -- (577,50.6) ;
\draw    (577,50.6) -- (577,110.3) ;

\draw (47,12) node [anchor=north west][inner sep=0.75pt]   [align=left] {\begin{minipage}[lt]{66.96pt}\setlength\topsep{0pt}
\begin{center}
{\scriptsize {\fontfamily{ptm}\selectfont Select }$\displaystyle o$ {\fontfamily{ptm}\selectfont obso }}\\{\scriptsize {\fontfamily{ptm}\selectfont cases and }$\displaystyle u$ }\\{\fontfamily{ptm}\selectfont {\scriptsize non-obso }}\\{\fontfamily{ptm}\selectfont {\scriptsize cases}}
\end{center}

\end{minipage}};
\draw (178,12) node [anchor=north west][inner sep=0.75pt]   [align=left] {\begin{minipage}[lt]{67.59pt}\setlength\topsep{0pt}
\begin{center}
{\scriptsize {\fontfamily{ptm}\selectfont Collect all the}}\\{\scriptsize {\fontfamily{ptm}\selectfont data available}}\\{\fontfamily{ptm}\selectfont {\scriptsize for the cases }}\\{\fontfamily{ptm}\selectfont {\scriptsize $\displaystyle o+u$}}
\end{center}

\end{minipage}};
\draw (308,10) node [anchor=north west][inner sep=0.75pt]   [align=left] {\begin{minipage}[lt]{66.54pt}\setlength\topsep{0pt}
\begin{center}
{\fontfamily{ptm}\selectfont {\scriptsize Add features}}\\{\fontfamily{ptm}\selectfont {\scriptsize provided by}}\\{\fontfamily{ptm}\selectfont {\scriptsize experts through}}\\{\fontfamily{ptm}\selectfont {\scriptsize survey}}
\end{center}

\end{minipage}};
\draw (447,10) node [anchor=north west][inner sep=0.75pt]   [align=left] {\begin{minipage}[lt]{53.86pt}\setlength\topsep{0pt}
\begin{center}
{\fontfamily{ptm}\selectfont {\scriptsize Construct dataset }}\\{\fontfamily{ptm}\selectfont {\scriptsize through expert }}\\{\fontfamily{ptm}\selectfont {\scriptsize annotation and}}\\{\fontfamily{ptm}\selectfont {\scriptsize opinion}}
\end{center}

\end{minipage}};
\draw (62,140) node [anchor=north west][inner sep=0.75pt]   [align=left] {\begin{minipage}[lt]{48.52pt}\setlength\topsep{0pt}
\begin{center}
{\fontfamily{ptm}\selectfont {\scriptsize Dimensionality }}\\{\fontfamily{ptm}\selectfont {\scriptsize reduction }}\\{\fontfamily{ptm}\selectfont {\scriptsize (PC and PCA)}}
\end{center}

\end{minipage}};
\draw (180,140) node [anchor=north west][inner sep=0.75pt]   [align=left] {\begin{minipage}[lt]{61.99pt}\setlength\topsep{0pt}
\begin{center}
{\fontfamily{ptm}\selectfont {\scriptsize Supervised}}\\
{\fontfamily{ptm}\selectfont {\scriptsize Learning}}\\{\fontfamily{ptm}\selectfont {\scriptsize (DTC)}}
\end{center}

\end{minipage}};
\draw (440,130) node [anchor=north west][inner sep=0.75pt]   [align=left] {\begin{minipage}[lt]{61.99pt}\setlength\topsep{0pt}
\begin{center}
{\fontfamily{ptm}\selectfont {\scriptsize Comparison of}}\\{\fontfamily{ptm}\selectfont {\scriptsize final results}}\\{\fontfamily{ptm}\selectfont {\scriptsize with results}}\\{\fontfamily{ptm}\selectfont {\scriptsize of expert opinion}}
\end{center}

\end{minipage}};
\draw (302,130) node [anchor=north west][inner sep=0.75pt]   [align=left] {\begin{minipage}[lt]{70.32pt}\setlength\topsep{0pt}
\begin{center}
{\fontfamily{ptm}\selectfont {\scriptsize Result analysis}}\\{\fontfamily{ptm}\selectfont {\scriptsize leading to decision}}\\{\fontfamily{ptm}\selectfont {\scriptsize making features}}\\{\fontfamily{ptm}\selectfont {\scriptsize identification}}
\end{center}

\end{minipage}};

\end{tikzpicture}
    \caption{Methodological Approach Process}
    \label{fig:methodological_approach_process}
\end{figure}

The section starts off by tackling the dataset formation methodology from the process described in Figure \ref{fig:methodological_approach_process} (shown in blue), afterwards the models used are discussed in Section \ref{sec:model_methodology} (shown in red), and finally the analysis method is tackled in Section \ref{sec:analysis_methodology} (shown in green).
\subsection{Dataset Formation Methodology}\label{sec:dataset_formation_methodology}
\subsubsection{Data Acquisition}\label{sec:data_acquisition}
A survey has been conducted regarding the features affecting obsolescence management solution selection of electrical products, e.g., blinkers, power supplies, etc. The survey was sent out to the obsolescence management team of SNCF RESEAU, who monitor, detect, and decide on the appropriate resolution for each obsolescence case. These personnel come from various backgrounds and domains of work, including the supply chain, technical management, technicians, engineering, and purchasing departments. For each obsolescence case, a set of features is recorded, but the knowledge is held by the experts of the team.

The dataset was constructed from numerous internal databases within the company following the formation of the set of features obtained from the survey conducted, the literature, and an internal company data visualization interface. The quality of the data varies depending on the database source. The data regarding the features referring to the technical properties of the various units and components are not always digitized, and thus the knowledge of the personnel as well as the archives of the company was accessed to be filled.
\subsubsection{Dataset Formalization}\label{sec:dataset_formalization}
The dataset is formalized in what follows. Given the dataset of obsolescence and non-obsolescence cases $\mathcal{D}$ such that $\mathcal{D}=\{(x_i,y_i)\in\mathcal{X}\times\mathcal{Y}|i=1,...,N_{o+u}\}$, where each case $x_i$ is a $D$-dimensional vector comprised of $D$ features $x_i\in\mathbb{R}^D$, and each class label, i.e., obsolescence management solution (e.g. replacement, repairing, etc. \cite{rust2022literature}), is represented by an integer $y_i\in\mathcal{Y}\equiv\{1,...,N\}$ resulting in $N$ distinct classes. Let $\mathcal{X}\equiv\mathbb{R}^D$ for generality. The set $\mathcal{X}$ is divided into two subsets: the obsolete set containing obsolete units, and the non-obsolete set containing non-obsolete units. The obsolete set is given by $\mathcal{O}=\{(x_i^o,y_i^o)\in\mathcal{X}\times\mathcal{Y}|i=1,...,N_o\}$, and the non-obsolete set is given by $\mathcal{U}=\{(x_i^u,y_i^u)\in\mathcal{X}\times\mathcal{Y}|i=1,...,N_u\}$. The obsolete and non-obsolete sets are disjoint $\mathcal{O}\cap\mathcal{U}=\emptyset$. Each of the sets above are represented by their corresponding matrices where $\mathcal{O}$, $\mathcal{U}$, and $\mathcal{X}$ are represented by the matrices $O\in\mathbb{R}^{o,D}$, $U\in\mathbb{R}^{u,D}$, and  $X\in\mathbb{R}^{o+u,D}$ respectively such that $X=\begin{pmatrix}
    O \\ 
    U
\end{pmatrix}$

The dataset $\mathcal{D}$ is composed of $o=316$ obsolete instances, and $u=10$ non-obsolete instances. The instances are labeled with one of the following classes of obsolescence remediation: existing stock ($7$ instances), Last-time Buy Order ($30$ instances), minor redesign ($28$ instances), major redesign ($124$ instances), and substitution ($126$ instances). The labeling process was conducted through expert annotation.
\subsection{Model Methodology}\label{sec:model_methodology}
As a preliminary process, and in order to combat the restrictive nature of binary values, all the binary fields are normalized on a scale from $[1,5]$ following the normal distribution as follows $1+4\frac{(x-\Bar{x})}{\sigma}$ where $x$ is the row feature vector and $\sigma$ is the standard deviation.
\subsubsection{Step 1: PC.} Highly correlated features pertain to variables that exhibit a strong linear relationship with one another, i.e., they convey nearly identical information. Consequently, this phenomenon can give rise to a predicament referred to as multicollinearity~\cite{montgomery2021introduction}, wherein the independent impact of each variable on the target variable becomes arduous to ascertain. Thus, to address this problem, the Pearson Correlation (PC) coefficient is calculated for $X^T$~\cite{mei2022modeling}. Any vector with a correlation above an arbitrarily selected $\alpha$ is removed following the backward elimination method~\cite{kotu2014predictive} from the matrix $X^T$, i.e., to reduce redundancy, thus reducing its dimensionality $X\in\mathbb{R}^{o+u,h}$ such that $h\leq D$.
\subsubsection{Step 2: PCA.} When confronted with a large number of features and a small dataset, the model may inadvertently fit the noise within the data rather than discerning the underlying patterns. Consequently, the ability of the model to generalize effectively diminishes~\cite{goodfellow2016deep}. By reducing the number of features the complexity of the model diminishes as well, thereby reducing the likelihood of overfitting the training data which leads to better generalization and higher accuracy on novel data~\cite{sammut2011encyclopedia}. Ergo, the Principal Component Analysis (PCA) algorithm is used which computes the orthonormal $\ell$-dimensional subspace $X^{*}\in\mathbb{R}^{o+u,\ell}$ of $X$ forming $\mathcal{X}^*$, and by consequence $\mathcal{D}^{*}$ such that $\ell<h$.
\subsubsection{Step 3: DTC.} The next and final step is to harness the knowledge that the experts possess through learning a function $f$ that takes in the expert provided instances and maps them to the expert classified class labels $f:\mathcal{X}^{*}\longrightarrow\mathcal{Y}$. The aim of this step is to mimic the reasoning behind the decisions made by the experts. For the reason of ease of interpretability, the Decision Tree Classifier (DTC) algorithm is used to learn the decision conditions of each class $y_i\in\mathcal{Y}$, i.e., obsolescence management solution.
\subsection{Analysis Methodology}\label{sec:analysis_methodology}
The first step of the analysis involves analyzing the mean decrease in Gini impurity within the learned DTC model $f$, which can be used to determine the importance of each feature: the higher the mean decrease in impurity, the more important the feature is considered~\cite{sandri2010analysis}. The second step is to determine the features that contribute the most to the formation of the orthonormal subspace $X^*$. This is done by taking the sum of the highest $\ell$ elements of the vectors of the column space of $X^*$, where each vector corresponds to a feature present in $X$~\cite{goodfellow2016deep}.
\section{Experimentation and Discussion}\label{sec:experimentation}
The number of initial features $D=16$ was reduced to $\ell=7$ principal components. The correlation threshold, set at $\alpha=0.15$, was determined by identifying the optimal value within the range $\alpha\in[-1,1]$ that maximizes the average accuracy of the DCT model while minimizing the dimensionality of $\mathcal{D}$, resulting in $D=12$. Similarly, the number of principal components $\ell=7$ was determined using a similar optimization method and objective. The optimal parameters were derived by selecting the values corresponding to the model with the highest accuracy among the $500$ tests conducted.

\begin{figure}[htbp]
    \centering
    \begin{subfigure}[b]{0.49\textwidth}
        \centering
        \includegraphics[width=\textwidth]{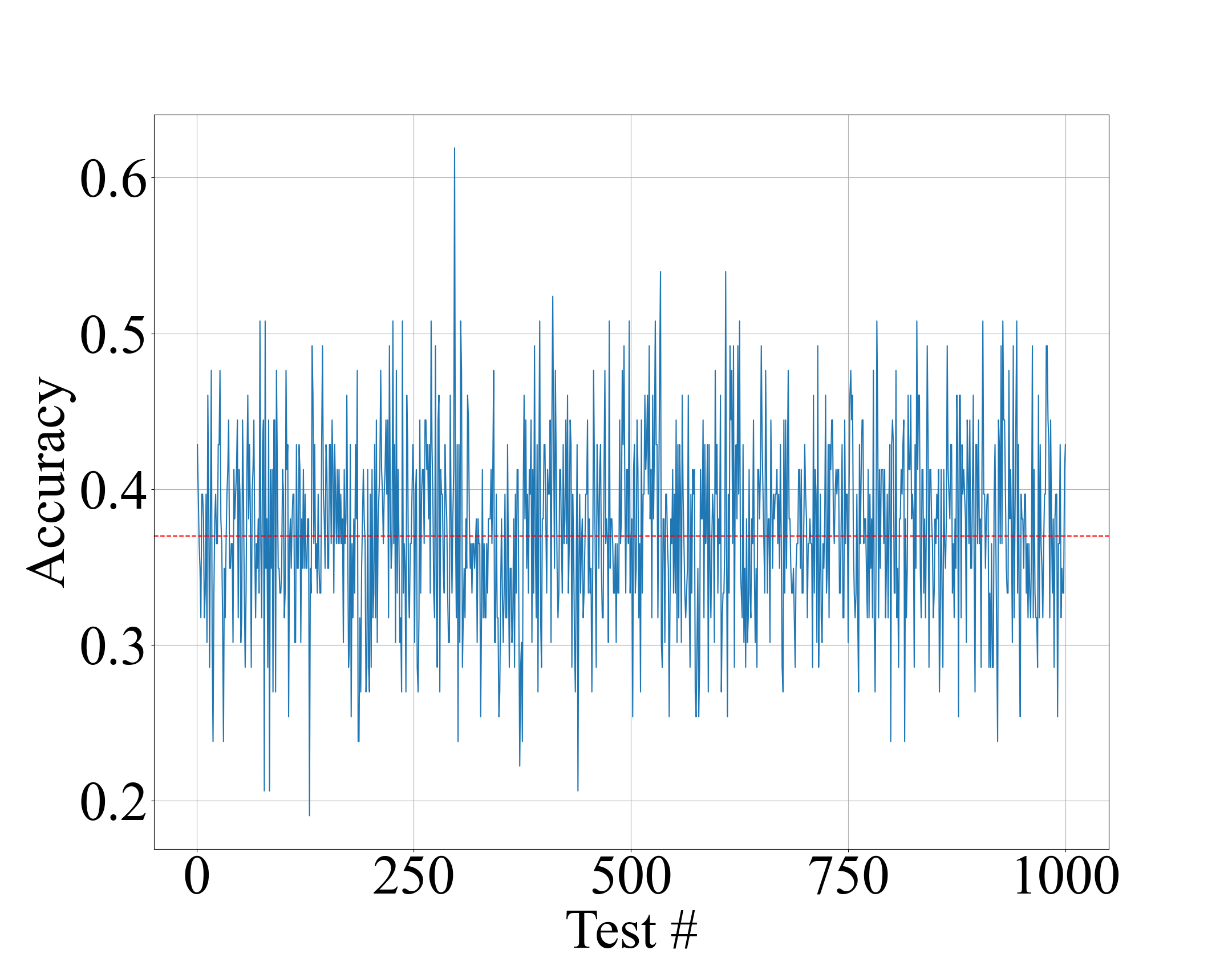}
        \caption{Model accuracy measurements.}
        \label{fig:accuracies_plot}
    \end{subfigure}
    \hfill
    \begin{subfigure}[b]{0.49\textwidth}
        \centering
        \includegraphics[width=\textwidth]{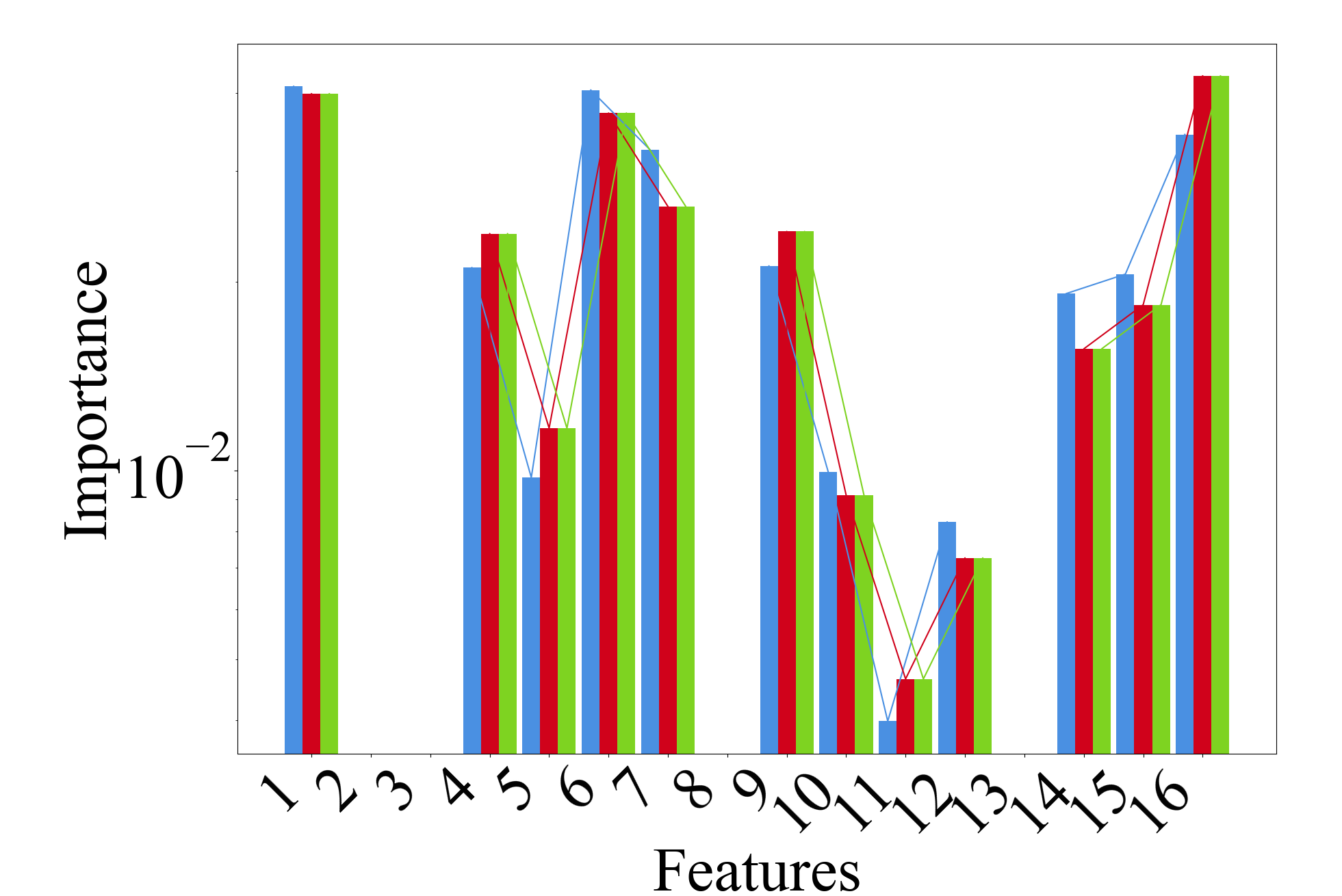}
        \caption{Feature contributions.}
        \label{fig:contributions_plot}
    \end{subfigure}
    \caption{Obtained results displaying feature contributions on a logarithmic scale, following the methodology outlined in Section \ref{sec:analysis_methodology}. Blue bars indicate mean contributions, red bars show values from the best model, and green bars represent contributions from the top-performing model on non-obsolete instances.
    }\label{fig:results}
\end{figure}

Afterwards, an additional $1,000$ tests were conducted on the proposed model using the obsolete set $\mathcal{O}$, which is shuffled before every test to reduce bias and evaluate its generalization. The maximum and minimum accuracy values of DCT are reported as $61.90\%$ and $19.05\%$, respectively. The standard deviation, arithmetic mean, and geometric mean of the accuracy measurements are $5.95\%$, $37.48\%$, and $37.00\%$, respectively. These values suggest a balanced spread, indicating that the accuracy measurements are not skewed and are symmetric. This implies a degree of inconsistency in the model's behavior, as illustrated in Figure \ref{fig:accuracies_plot}, but it is not considered erratic. The top-performing model attained a $40\%$ accuracy on the non-obsolete set $\mathcal{U}$ suggesting a potential application for monitoring using the model.

The sorted list of highest average feature contributions aligns closely with the model achieving the highest accuracy on both $\mathcal{O}$ and $\mathcal{U}$. In both lists, the top three elements include ``repair complexity,'' ``several components are involved,'' and ``supplied as new, no longer supply, no longer newly supply, reparable only.'' However, it contradicts the list derived from the combined options of experts ``existing stock quantity,'' ``existing substitution product,'' and ``security block/product''), highlighting a lack of consensus on the subject. For detailed lists, refer to the GitHub repository: \url{https://github.com/Inars/A-Hybrid-Approach-for-Identifying-Obsolescence-Features}.

The results shown display a discrepancy within the decision making process of the SNCF RESEAU obsolescence management team. They emphasize the need of a unified decision making model that attempts to deduce consensus by optimally finding a resolution to a given obsolescence case using the features that are most relevant.
\section{Conclusion and Perspective}\label{sec:conclusion_and_perspective}
This paper presents a hybrid methodology designed to identify feature contributions that influence decision-making in obsolescence management resolutions. The proposed method highlights the divergence of expert opinions on the subject. During the dataset formation process, a noticeable discrepancy in the number of class instances was observed, necessitating a substantial increase in instances. Two key research areas emerge: data generation techniques to address data scarcity and k-shot learning models for resolving class imbalances.
%
%
\bibliographystyle{apalike}
\bibliography{references}

\begin{thebibliography}{}

\bibitem[Goodfellow et~al., 2016]{goodfellow2016deep}
Goodfellow, I., Bengio, Y., and Courville, A. (2016).
\newblock {\em Deep Learning}.
\newblock MIT Press.

\bibitem[Kotu and Deshpande, 2014]{kotu2014predictive}
Kotu, V. and Deshpande, B. (2014).
\newblock {\em Predictive analytics and data mining: concepts and practice with rapidminer}.
\newblock Morgan Kaufmann.

\bibitem[Mei et~al., 2022]{mei2022modeling}
Mei, K., Tan, M., Yang, Z., and Shi, S. (2022).
\newblock Modeling of feature selection based on random forest algorithm and pearson correlation coefficient.
\newblock In {\em Journal of Physics: Conference Series}, volume 2219, page 012046. IOP Publishing.

\bibitem[Montgomery et~al., 2021]{montgomery2021introduction}
Montgomery, D.~C., Peck, E.~A., and Vining, G.~G. (2021).
\newblock {\em Introduction to linear regression analysis}.
\newblock John Wiley \& Sons.

\bibitem[Pingle, 2015]{pingle2015selection}
Pingle, P. (2015).
\newblock {\em Selection of obsolescence resolution strategy based on a multi criteria decision model}.
\newblock PhD thesis, Iowa State University.

\bibitem[Porter, 1998]{porter1998economic}
Porter, G.~Z. (1998).
\newblock {\em An economic method for evaluating electronic component obsolescence solutions}.
\newblock Boeing.

\bibitem[Renaud et~al., 2018]{renaudmacup}
Renaud, G., Abiven, Y.-M., Ta, F., Tran, Q., and Zhang, S. (2018).
\newblock {MACUP} ({M}aterial for data {AC}quisition {-} {UP}grade): {P}roject {F}ocusing on {DAQ} {H}ardware {A}rchitecture {U}pgrades for {SOLEIL}.
\newblock In {\em Proc. of International Conference on Accelerator and Large Experimental Control Systems (ICALEPCS'17), Barcelona, Spain, 8-13 October 2017}, pages 1330--1334.

\bibitem[Rust et~al., 2022]{rust2022literature}
Rust, R.~M., Elshennawy, A., and Rabelo, L. (2022).
\newblock A literature review on mitigation strategies for electrical component obsolescence in military-based systems.
\newblock {\em South African Journal of Industrial Engineering}, 33(1):25--38.

\bibitem[Sammut and Webb, 2011]{sammut2011encyclopedia}
Sammut, C. and Webb, G.~I. (2011).
\newblock {\em Encyclopedia of machine learning}.
\newblock Springer Science \& Business Media.

\bibitem[Sandri and Zuccolotto, 2010]{sandri2010analysis}
Sandri, M. and Zuccolotto, P. (2010).
\newblock Analysis and correction of bias in total decrease in node impurity measures for tree-based algorithms.
\newblock {\em Statistics and Computing}, 20:393--407.

\bibitem[Singh and Sandborn, 2006]{singh2006obsolescence}
Singh, P. and Sandborn, P. (2006).
\newblock Obsolescence driven design refresh planning for sustainment-dominated systems.
\newblock {\em The Engineering Economist}, 51(2):115--139.

\bibitem[Zheng et~al., 2015]{zheng2015design}
Zheng, L., Terpenny, J., and Sandborn, P. (2015).
\newblock Design refresh planning models for managing obsolescence.
\newblock {\em Iie Transactions}, 47(12):1407--1423.

\bibitem[Zolghadri et~al., 2021]{zolghadri2021obsolescence}
Zolghadri, M., Addouche, S.-A., Baron, C., Soltan, A., and Boissie, K. (2021).
\newblock Obsolescence, rarefaction and their propagation.
\newblock {\em Research in Engineering Design}, 32(4):451--468.

\end{thebibliography}
\end{document}